\documentclass[aps,prd,preprint,nofootinbib]{revtex4}
\usepackage{amsfonts}
\usepackage{mathrsfs}
\usepackage{graphicx}
\usepackage{amsmath}
\usepackage{amssymb}
\usepackage{bm,multirow}
\usepackage{multirow}
\usepackage{epsfig}
\usepackage{subfig}
\newcommand{\bqa}{\begin{eqnarray}}
\newcommand{\eqa}{\end{eqnarray}}
\newcommand{\beq}{\begin{equation}}
\newcommand{\eeq}{\end{equation}}

\graphicspath{{fig/}{dia/}} \DeclareGraphicsExtensions{.eps}
\usepackage{subfig}
\begin{document}
\baselineskip 20pt

\title{X(16.7) Production in Electron-Position Collision}

\author{Jun Jiang$^{a}$\footnote{jiangjun13b@mails.ucas.ac.cn}}
\author{Long-Bin Chen$^{b}$\footnote{chenglogbin10@mails.ucas.ac.cn}}
\author{Yi Liang$^{a}$\footnote{alvan@ucas.ac.cn}}
\author{Cong-Feng Qiao$^{a,c}$\footnote{qiaocf@ucas.ac.cn, corresponding author}}

\affiliation{$^a$School of Physical Sciences, University of Chinese Academy of Sciences, YuQuan Road 19A, Beijing 100049, China\\
$^b$School of Physics \& Electronic Engineering, Guangzhou University, Guangzhou 510006, China\\
$^c$CAS Center for Excellence in Particle Physics, Beijing 100049, China}

\author{~\\}

\begin{abstract}

The anomaly found in the excited $^8\text{Be}$ nuclear transition to its ground state is attributed to a spin-1 gauge boson $X(16.7)$. To hunt for this boson, we propose two traps: $e^+e^-\to X\gamma$ and $J/\psi\to X\gamma$, both following with $X\to e^+e^-$ decay. We adopt the ``vector minus axial-vector'' interaction hypothesis. Analysis on the $X(16.7)$ decay length, production rates, differential distribution with respect to the $e^+e^-$ invariant-mass spectrum, and the signal-to-noise ratios (SNR) after the smearing at BESIII detector are discussed in detail. Given the coupling strength of $X$ to vector/axial-vector currents $g_f^{v/a}\sim10^{-3}$ at BESIII: (1) there would be about $6000$ $X$ measurable events per year in electron-positron collision, yet with a large background after smearing; (2) while in $J/\psi$ decays, we find that the axial-vector current may come into play; though merely 52 events may appear, the SNR are inspiring even after smearing.

\vspace {7mm} \noindent {PACS numbers: 14.70.Pw, 12.60.Cn, 13.66.Hk, 13.20.Gd}

\end{abstract}
\maketitle

\section{Introduction}

So far, the Standard Model (SM) has achieved a great success in experiencing numerous experimental tests \cite{Patrignani:2016xqp}. And the search of new physics beyond SM is now a major activity for both experimental and theoretical physicists. Gauge bosons, $\gamma,~Z^0,~W^{\pm}$ and gluons, as the messengers of the SM interaction forces, play crucial role in revealing the nature of interactions. A new gauge boson (or the fifth force), then becomes the prey which we all are hunting for.

Finding out whether there exists a new type of force beyond SM is a very tempting task. The ATLAS and CMS Collaborations at LHC searched for the TeV $Z^{\prime}$ boson using the Run 2 data \cite{Aad:2015osa,CMS:2016zxk,ATLAS:2016cyf,CMS:2016abv,Policicchio:2016evl}. For B-factories, the BaBar Collaboration explored the 0.02$\sim$10.2 GeV region using the $e^+e^-$ invariant-mass spectrum and upper limits on mixing strength between the dark photon (U) and SM photon ($\gamma$) were placed \cite{Lees:2014xha}. At BESIII detector, the 1.5$\sim$3.4 GeV mass range was explored and limits on the U$-\gamma$ mixing strength were also set \cite{Ablikim:2017aab}. Similarly in the KLOE-2 experiment, they searched for the dark photon in both $e^+e^- \to \text{U}\gamma$ \cite{Anastasi:2015qla} and $\phi \to \text{U}\eta$ processes \cite{Babusci:2012cr}. The NA48/2 experiment at CERN also searched for the dark photon in $\pi^0$ decays \cite{Batley:2015lha}. While in HADES experiment, the dark photon search was carried out using $e^+e^-$ spectrum in the p-p, N-b reactions, as well as the Ar + KCl reaction \cite{Agakishiev:2013fwl}. These new and dark bosons, $Z^{\prime}$ and U, were also probed with the precise electroweak data \cite{Erler:2009jh}, in neutrino-electron scattering experiments \cite{Bilmis:2015lja}, planned and future experiments \cite{Ilten:2015hya,Wojtsekhowski:2017ijn,Corliss:2017tms,Gninenko:2017acc}, \emph{etc.}, and have attracted the bright physicists \cite{Alexander:2016aln,Kozhuharov:2016tdb,Denig:2016dgi}.

In 2016, an extraordinary experimental phenomenon was observed in the isoscalar $^8$Be nuclear transition, $^8\text{Be}^*$ $\rightarrow$ $^8\text{Be}$ \cite{Krasznahorkay:2015iga}, and new measurements are presented three times recently \cite{Krasznahorkay:2017bwh,Krasznahorkay:2017gwn,Krasznahorkay:2017qfd}. A significant enhancement relative to the internal pair creation was observed at large angles in the invariant-mass distribution of electron-positron pairs production. However, no anomaly is seen in the isovector $^8\text{Be}^{*\prime}$ $\rightarrow$ $^8\text{Be}$ transition. This observation is hard to be understood within the regime of conventional theory, but could be attributed to a neutral isoscalar particle $X$ with the mass of 16.7 MeV and the saturating decay $X \rightarrow  e^+ e^-$ beyond SM.

At the beginning, this anomaly has been interpreted as a new vector boson which mediates a weak fifth force beyond SM \cite{Feng:2016jff}, and a realistic model for the fifth force is also proposed \cite{Gu:2016ege}. Other possible explanations on this new boson, light pseudoscalar boson, protophobic vector boson, axial-vector particle, \emph{etc.}, are also widely studied \cite{Ellwanger:2016wfe,Feng:2016ysn,Kozaczuk:2016nma,Neves:2016ugb,DelleRose:2017phz,Neves:2017rcn,Fornal:2017msy,Zhang:2017zap}.
Among them, the axial-vector particle proposal is suggested in Ref.s \cite{Feng:2016jff,Feng:2016ysn} but not pursued systematically. In Ref. \cite{Kozaczuk:2016nma}, the production of a hidden vector boson with axial-vector couplings to leptons and light quarks in the isoscalar $^8\text{Be}^*$ $\rightarrow$ $^8\text{Be}$ nuclear transition is investigated. Note, along with the axial-vector couplings to the Standard Model fermions, we need to devote some effort to obtain a UV-complete anomaly-free theory \cite{Kahn:2016vjr,Ismail:2016tod}. Besides, models including this $X$ boson as a mediator to the dark sector or giving constrains on dark matters are discussed \cite{Kitahara:2016zyb,Jia:2016uxs,Chen:2016tdz,Yamamoto:2017ypv}. This boson is also introduced to account for some other anomaly observations \cite{Liang:2016ffe,Jia:2017iyc} or as the light gauge boson mediator in the rare kaon and pion decays \cite{Chen:2016kxw,Chiang:2016cyf}. Meanwhile, suggestions for future experimental research are proposed \cite{Raggi:2014zpa,Araki:2017wyg,Kozaczuk:2017per,Alikhanov:2017cpy,Kozhuharov:2017qjo}.

Scientifically, to investigate further in experiments and get more knowledge of this-yet-not-independently-verified particle is currently the most important task among all the studies. In this work, we estimate the production of this $X$ boson associated with a photon in electron-positron collision, as well as in the $J/\psi$ decays. The complete ``vector minus axial-vector (V-A)'' interaction is considered seriously, and we find that the axial-vector current contribution plays an important role, especially in $J/\psi$ decays. Various analysis on its production are presented in detail, particularly the invariant-mass spectrum of the electron-positron pairs in final states, and the signal-to-noise ratios after the ``smearing''.

\section{Traps Arranged for $X(16.7)$}

To be a new member of the Particle Zoo, $X(16.7)$ needs to be tagged. Since the proposals of scalar bosons are excluded \cite{Feng:2016jff}, we here focus on the spin-1 gauge boson hypothesis. And the Lagrangian added to the Standard Model one can be formulated as
\bqa
\mathcal{L}_X=-\frac{1}{4}X_{\mu\nu}X^{\mu\nu} + \frac{1}{2}m_X^2X_{\mu}X^{\mu}-\sum_f e\bar{f}\gamma_\mu (g^v_f-g^a_f\gamma_5) f X^\mu.\label{lag}
\eqa
Here, $e$ is the electron charge and $g^{v/a}_f$ denote the coupling strength of $X$ to vector/axial-vector currents, which means $X$ boson can either be a massive $\gamma$-like particle or a $Z^0$-like one. For the $g^{v}_e$, it has been constrained by experimental data to the region of $2\times 10^{-4} \leq|g^v_e| \leq1.4\times 10^{-3}$ \cite{Feng:2016jff}. In this paper, we will take the ``vector minus axial-vector (V-A)'' interaction vertex as a general situation, one can obtain the vector one readily by taking the coupling parameter $g^a_f=0$.

\subsection{$e^+e^- \to X+\gamma$ with $X \to e^+e^-$}

\begin{figure}[t]
\begin{center}
\includegraphics[scale=0.8]{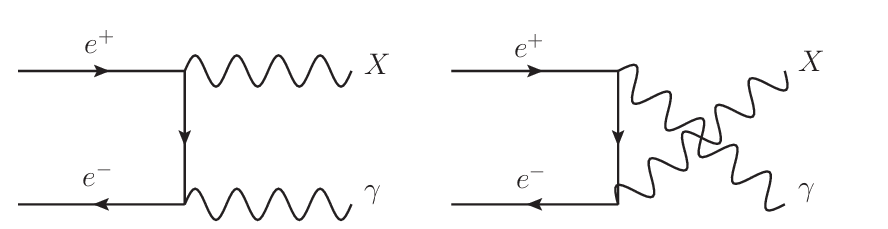}
\caption{The Feynman diagrams which contribute to the $X$ boson production associated with a photon in electron-positron collision.}\label{fxp}
\end{center}
\end{figure}

\begin{figure}[t]
\begin{center}
\includegraphics[scale=0.8]{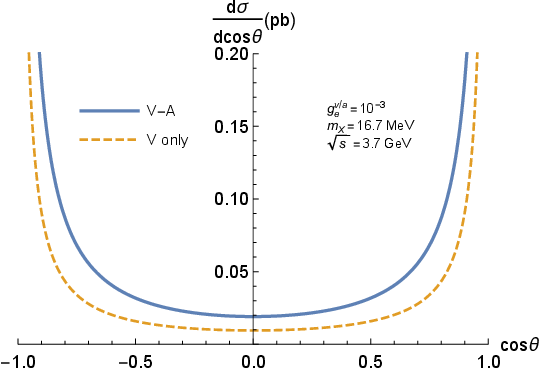}
\includegraphics[scale=0.85]{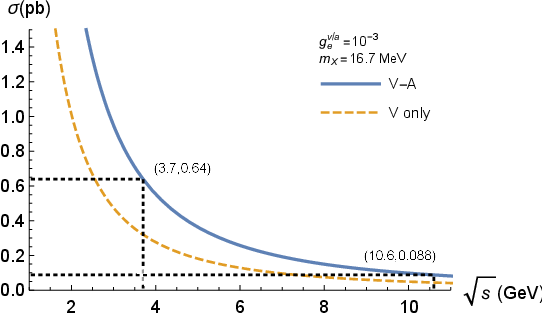}
\caption{Left: the differential cross-sections with respect to $\cos\theta$. Right: the cross-section as a function of the CMS energy $\sqrt{s}$. The legend ``V-A'' refers to the complete ``vector minus axial-vector'' contributions, while ``V only'' stands for the vector case only.}\label{coss}
\end{center}
\end{figure}

Since the mass of $X$ boson $m_X$ is much smaller than the energy of usual electron-positron colliders, its production always associates with another gauge boson radiation, \emph{i.e.} the photon $\gamma$, as shown in Fig. \ref{fxp}.
The differential cross-section of the process can be readily obtained,
\bqa
\frac{d\sigma}{d\cos\theta}=\frac{2\pi \alpha^2(s-m_X^2)}{16s\sqrt{s(s-4m_e^2)}}\times\big((g^v_e)^2 |M_v|^2+(g^a_e)^2 |M_a|^2\big), \label{dif-eq1}
\eqa
\bqa
|M_v|^2&=
\frac{32 s (\cos^2\theta (4 m_e^2-s) (s (4 m_e^2+s)+m_X^4)+s (-16m_e^4+4 m_e^2 (s-2 m_X^2)+m_X^4+s^2))}{(m_X^2-s)^2 (\cos^2\theta (4 m_e^2-s)+s)^2}-16, \nonumber\\
|M_a|^2&=
\frac{256 m_e^2 s^2 (4 m_e^2-m_X^2)}{(m_X^2-s)^2 (\cos^2\theta (4 m_e^2-s)+s)^2}-\frac{32 s (2 m_e^2 (m_X^4-6 m_X^2 s+s^2)+m_X^2 (m_X^4+s^2))}{(m_X^3-m_X s)^2 (\cos^2\theta (s-4 m_e^2)-s)}-16. \nonumber
\eqa
In which $\theta$ is the emitting angle of photon with respect to the $e^+e^-$ beam axis. Taking the fine structure constant $\alpha={1}/{137}$ and center-of-mass energy (CMS) $\sqrt{s}=3.7$ GeV, we can obtain the differential cross-section as being displayed in Fig. \ref{coss} (left). One may notice from the figure that the main contribution comes from the region where $|\cos\theta|$ is large. Taking the high-energy limit ($ \sqrt{s} \gg  m_X, m_e$), the differential cross-section turns to
\bqa
\frac{d\sigma} {d\cos\theta} = \frac{2\pi\alpha^2} {s\sin^2\theta}\big((g^v_e)^2(\cos^2\theta+1)+(g^a_e)^2(\cos^2\theta+5)\big),
\label{dif-eq2}
\eqa
which agrees with the result for $e^+e^-\rightarrow2\gamma$ by taking $g^v_e=1,~g^a_e=0$ \cite{peskin}.

Integrating over $\cos\theta$ of the differential cross-section Eq. (\ref{dif-eq1}), one can get the colliding energy dependence of the cross-section, which is presented in Fig. \ref{coss} (right). Where two CMS energies ($\sqrt{s}=3.7,10.6$ GeV) are highlighted. The first one is the typical CMS energy of BESIII detector, which has the luminosity of $10^{33}$ cm$^{-2}$s$^{-1}$ at $\sqrt{s}=3.7$ GeV. While the second one is that of B-factories, \emph{i.e.} the BaBar and Belle Collaborations. We find that the cross-section drops by about one order of magnitude when CMS energy increases from 3.7 GeV to 10.6 GeV. Another important conclusion drawn from Fig. \ref{coss} is that the contribution of axial-vector current (``A only'') and the vector current part (``V only'') have almost the same share when taking $g^v_e = g^a_e$. However, the ``A only'' contribution has not been taken into account in the literature seriously.

At the CMS energy of $\sqrt{s}=3.7$ GeV, and taking $g^{v/a}_e = 10^{-4}\sim10^{-3}$, we find the cross-section of $e^+e^- \to X+\gamma$ lies in the region of $0.0064\sim0.64$ pb, which happens to be two times the amount of the ``V only'' result. Since the BESIII detector can cover about 93\% of the 4$\pi$ solid angle, and with the luminosity of $L\sim10^{4}$ pb$^{-1}$year$^{-1}$, there will be about $60 \sim 6000$ $X$ bosons to be produced per year and more in yet collected data.\footnote{Roughly, a collider operates only about $10^{7}$ s in one year \cite{Han:2005mu}. So it is customary to estimate that $10^{33}$cm$^{-2}$s$^{-1}$$\simeq10^{4}$ pb$^{-1}$year$^{-1}$. Moreover, when evaluating the events, we simply multiply the cross-section with the integrated luminosity and the percentage of the solid angle coverage.} At Babar, with the copious $514~\text{fb}^{-1}$ data \cite{Lees:2014xha} and about 90\% solid angle coverage \cite{Aubert:2001tu}, the $X$ boson events would be around $4.1\times(10^2\sim10^4)$.

Experimentally, the $X$ boson would be reconstructed by its decay products. \emph{A priori} the $X$ boson can decay into $e^+e^-,~\nu\bar{\nu},~3\gamma$ or unknown particles, and $\nu\bar{\nu}$ and $3\gamma$ decay modes are highly suppressed \cite{Feng:2016jff}. So we assume that the $X$ boson decays to electron-positron pairs in saturation. And the decay width reads
\bqa
\Gamma(X\rightarrow e^+e^-)= \frac{\alpha\sqrt{m^2_X-4m_e^2}}{3m^2_X}\big((g^v_e)^2(m_X^2+2m_e^2)+(g^a_e)^2(m_X^2-4m_e^2)\big),
\eqa
which is consistent with the ``V only'' result by taking $g^a_e=0$ \cite{Pospelov:2008zw}. Since $m_X^2 \gg m_e^2$, the decay width is about two times the amount of that in ``V only'' situation when taking $g^v_e= g^a_e$. Given the order of magnitude of $g^{v/a}_e$ as $10^{-4}\sim10^{-3}$, the $X$ boson decay width varies as $~8.1\times(10^{-4}\sim10^{-2})$ eV, which corresponds to the lifetime $\tau = 8.1 \times(10^{-15}\sim10^{-13})~\text{s}$ at the $X$ boson rest frame. While in the experiment frame, the velocity of $X$ boson and the energy of the emitting photon reads $v = \frac{E_0} {\sqrt{E_0^2+m_X^2}}$ and $E_0=\frac{s-m_X^2}{2\sqrt{s}}$ respectively. After performing the Lorentz boost, the lifetime of $X$ boson could increase by about two orders of magnitude, \emph{i.e.}, $9.0 \times(10^{-13}\sim10^{-11})~\text{s}$. Hence, the decay length in the experiment frame would be
\beq
L = \frac{1.23\times10^4}{(g_e)^2}\times\frac{1}{\sqrt{1-v^2}} \times \hbar \times c,
\eeq
where $g_e^{v}= g_e^{a}$ is adopted and symbolized as $g_e$, $\hbar$ and $c$ are the reduced Planck constant and velocity of light respectively. Then numerically we have $0.27$ mm $ < L < 27$ \text{mm} at $\sqrt{s}=3.7$ GeV, which is measurable at BESIII. At the B-factories, whose CMS energy is about $10.6~\text{GeV}$, the decay length can reach $0.77~\text{mm} \sim 77~\text{mm}$. While at CLOE with $\sqrt{s}=1.0195$ GeV, the decay length of $X$ boson is about 10\% of that at B-factories. Evidently, it is more attainable to measure the decay length of $X$ boson at BaBar/Belle than at BESIII or CLOE. The measurement of the decay length is meaningful not only for the aim of disentangle signals from the background, but also for the determination of the coupling strength $g^{v/a}_e$.
\begin{figure}[t]
\begin{center}
\includegraphics[scale=0.5]{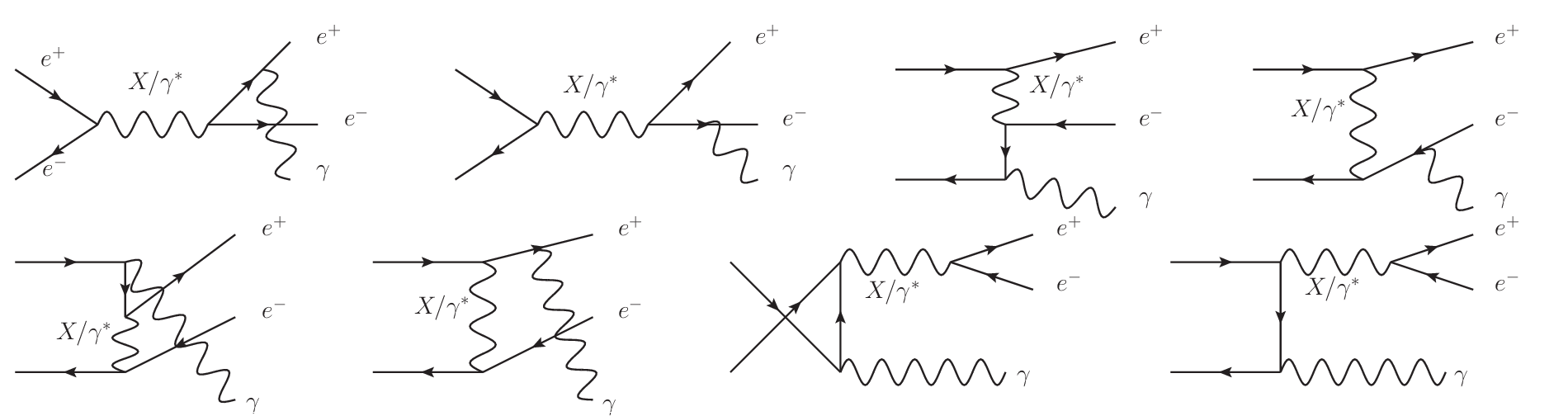}
\caption{The Feynman diagrmas for the signal/background processes $e^+e^- \xrightarrow[]{X/\gamma^*} e^+e^-\gamma$.}\label{eeg}
\end{center}
\end{figure}

It's time to set the trap with the $X \to e^+e^-$ decay mode. And the dominant background originates from the virtual photon propagated scattering process, $e^+e^- \xrightarrow[]{\gamma^*} e^+e^-\gamma$. We consider Feynman diagrams that contain both $X$ and $\gamma^*$ as inner propagators, which have been displayed in Fig. \ref{eeg}.

\begin{figure}[t]
\begin{center}
\includegraphics[scale=0.65]{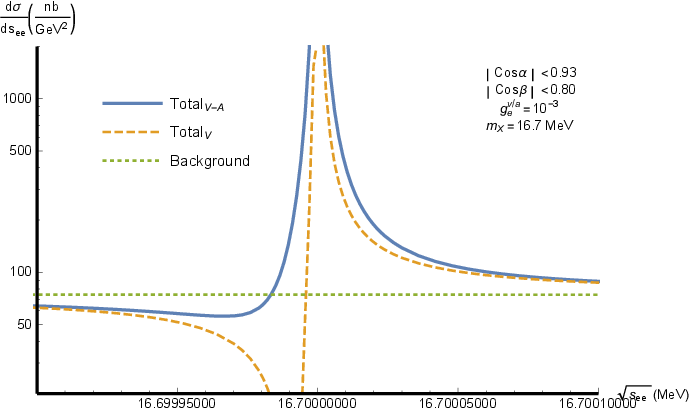}
\includegraphics[scale=0.65]{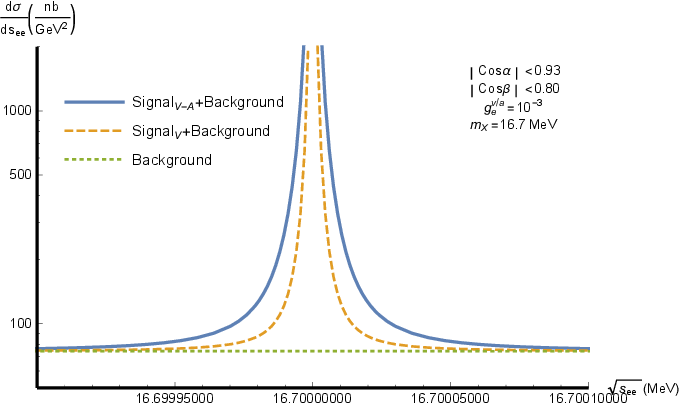}
\caption{The differential cross-section with respect to the invariant-mass distribution of $e^+e^-$ pairs for the $e^+e^- \xrightarrow[]{X/\gamma^*} e^+e^-\gamma$ processes, where we have adopted the good tracks conditions for both electron/positron and photon. Left: ``Total'' means it contains the  contribution of cross-terms between the $X$-propagated Feynman diagrams and the $\gamma^*$-propagated ones. Right: the contribution of cross-terms is excluded.}\label{mee}
\end{center}
\end{figure}

Here, we would evaluate the differential distribution with respect to the invariant-mass spectrum of electron-positron pairs in final states under the BESIII experiment conditions as a fine example. At BESIII, the good charged tracks are constrained in the region of $|\cos\alpha|<0.93$, while the photon selection condition is $|\cos\beta|<0.8$ with the energy $E>25~\text{MeV}$ for the barrel \cite{Ablikim:2009aa},\footnote{For the endcap, the constrain on photon is $0.84<|\cos\beta|<0.92$ with the energy $E>50~\text{MeV}$. Since the polar angle coverage of the endcap region is quite narrow, we do not take this part into account in this numerical evaluation.} here $\alpha/\beta$ are the polar angles with respect to the $e^+e^-$ beam axis. In Fig. \ref{mee} we show the invariant-mass ($\sqrt{s_{ee}}$ or $M_{ee}$) distribution of $e^+e^-$ pairs in final states for the differential cross-section, where the constrains on $\alpha/\beta$ are adopted. Given $g_e^{v/a}=10^{-3}$, we present the contribution of complete vector minus axial-vector current (``V-A'') and the vector current (``V only'') one separately. Here, the legend ``Background'' refers to the contribution of $e^+e^- \xrightarrow[]{\gamma^*} e^+e^-\gamma$ process; ``Signal'' stands for the contribution of $e^+e^- \xrightarrow[]{X} e^+e^-\gamma$ only; and the ``Total'' contains both the previous two parts and those from the cross-terms between them. By comparing the ``Total'' and ``Signal+Background'' figures in the left and right diagrams of Fig. \ref{mee}, one can easily obtain the running line-shape of the cross-terms.

From Fig. \ref{mee}, we can see that the $X$ boson acts as a rather sharp peak above the background around the mass of 16.7 MeV, which tells that a precise measurement on the invariant-mass spectrum of $e^+e^-$ pairs in final states will greatly suppress the background. Then, one may wonder if the energy resolution ($\delta_E$) of the BESIII detector is sufficient to undertake this invariant-mass analysis, \emph{i.e.}, the capability of identifying the signals over the background after ``smearing''. In the following, we take the sample of 6000 $X$ events as an example to illustrate the smearing and estimate the signal-to-noise ratios after smearing at BESIII.

\begin{figure}[t]
\begin{center}
\includegraphics[scale=0.25]{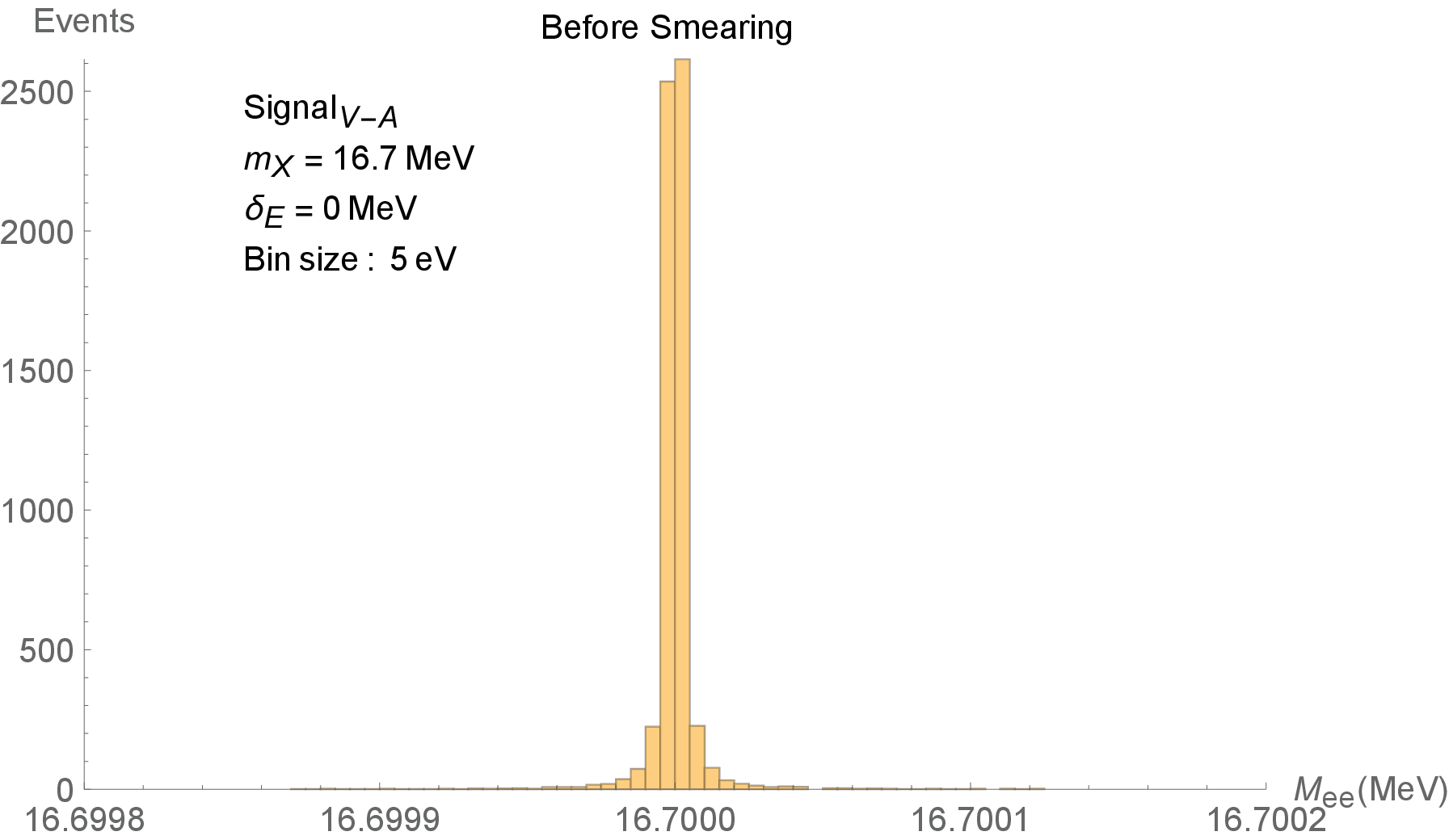}
\includegraphics[scale=0.25]{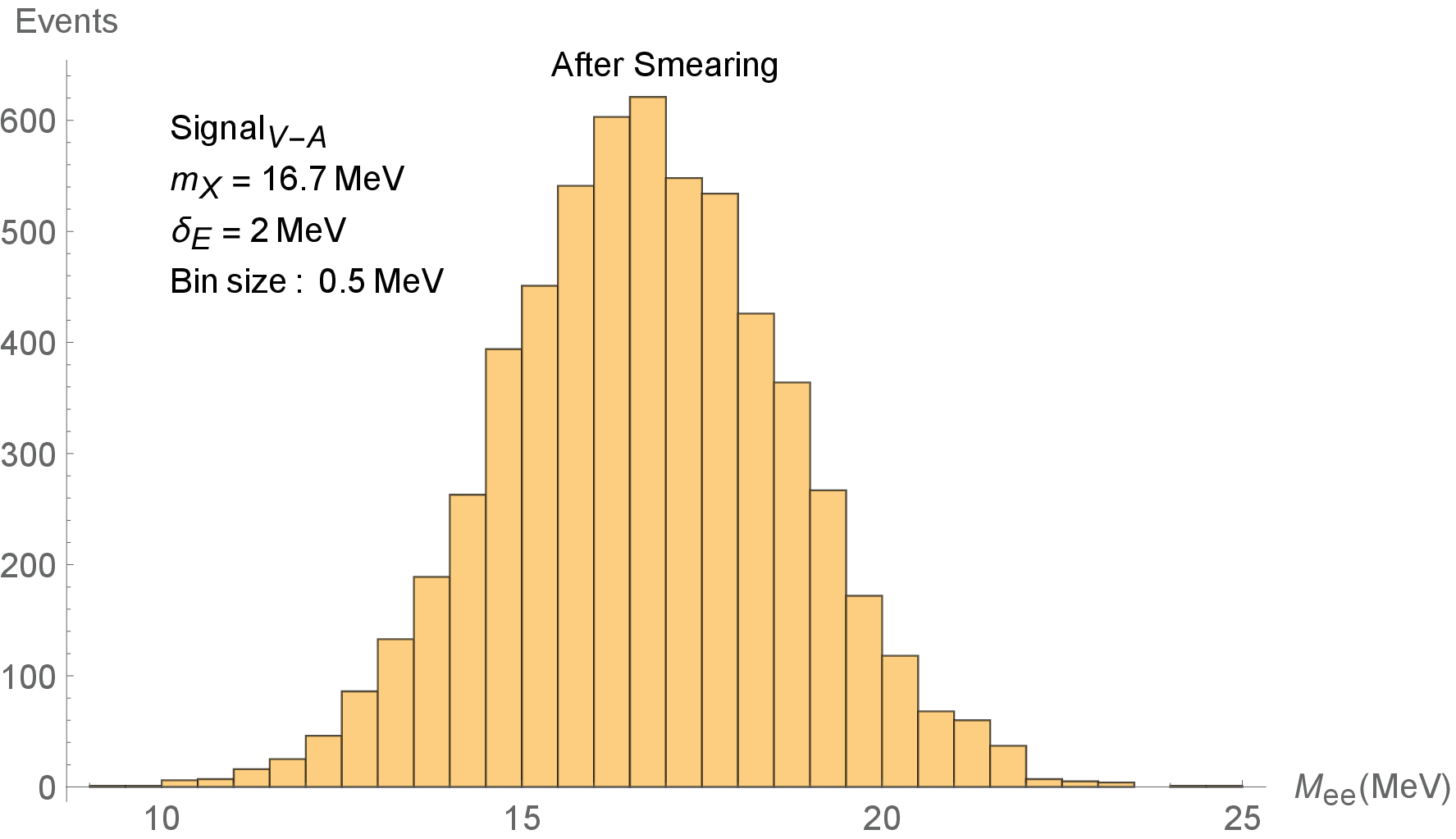}
\includegraphics[scale=0.25]{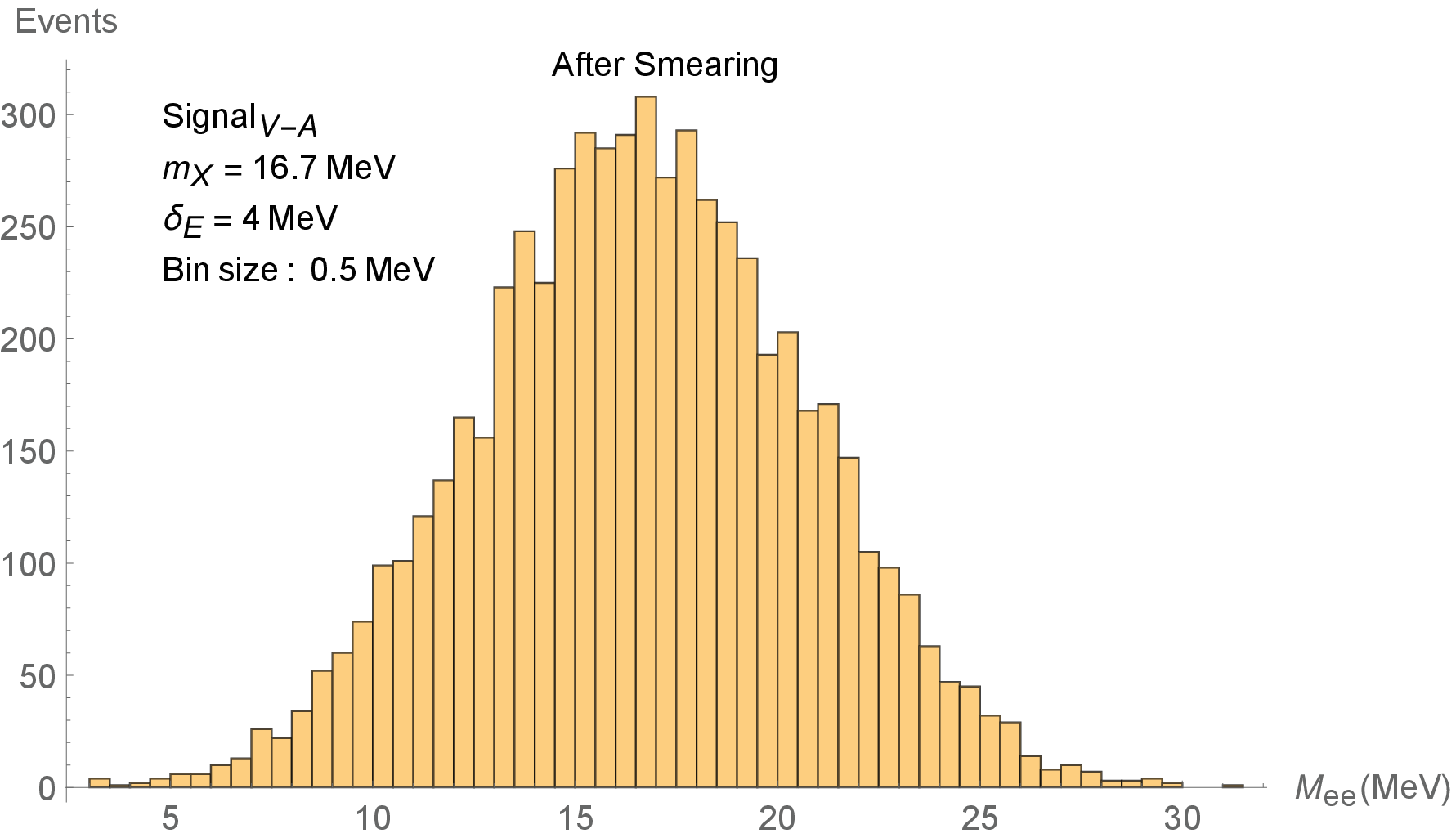}
\includegraphics[scale=0.25]{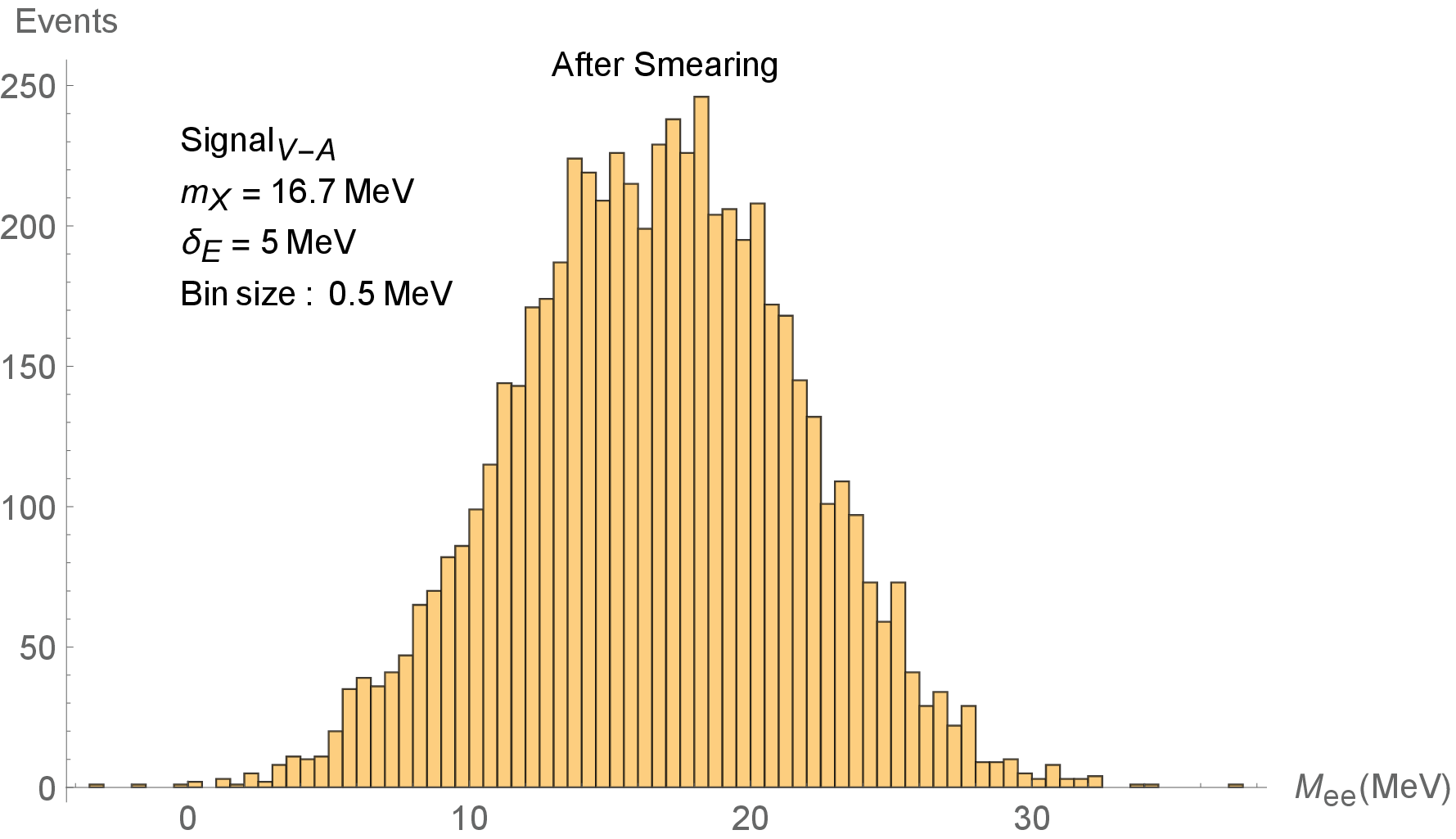}
\caption{The $X$ resonance bump of ``Signal$_{V-A}$'' before and after smearing simulated with 6000 $X$ events. The (b), (c) and (d) diagrams correspond to the energy resolution of $\delta_E=2,~4,~5$ MeV after smearing respectively.}\label{smear}
\end{center}
\end{figure}

Theoretically, these 6000 $X$ boson events would be reconstructed precisely at $M_{ee}=16.7$ MeV in the $e^+e^-$ invariant-mass spectrum, as is shown in Fig. \ref{smear} (a). However, experimentally the detector has the restricted energy resolution. So, some $X$ events would not be found at $M_{ee}=16.7$ MeV, but positioned at somewhere deviated from it. And the deviation amounts comply with a Gaussian distribution with the expectation of zero and the standard deviation of the energy resolution ($\delta_E$). In Fig. \ref{smear} (b, c, d), we present how the 6000 $X$ events are distributed in a wider invariant-mass range around $M_{ee}=16.7$ MeV in the spectrum after smearing, where $\delta_E=2,~4,~5$ MeV accordingly.

As is known, when the resonance peak is smeared to a wider $M_{ee}$ range, more background events would be included. To make it clear, we have considered four ranges of the $e^+e^-$ invariant-mass spectrum $M_{ee}$ around the $X$ resonance, and the corresponding signal-to-noise ratios (SNR) are presented in Tab. \ref{snr}. Note, for the signal events (S), we adopt the rough estimate of $X$ boson events produced in $e^+e^-\to X \gamma$, since the following decay $X \to e^+e^-$ is in saturation; while the background events (B) are obtained within the stringent constrains of the good tracks conditions. In Tab. \ref{snr}, we find that the SNR is quite encouraging in the first $M_{ee}$ range (near the $X$ boson peak). And the last three mass ranges are roughly corresponding to those with the energy resolution of $\delta_E=2,~4,~5$ MeV (Fig. \ref{smear} (b,~c,~d)) respectively, where the background is much more noisy. With the decay length of $X$ boson to further increase the SNR, we believe that these results are inspiring in the future $X$ boson hunting if $g_e^{v/a}\sim10^{-3}$.

In fact, the BaBar, BESIII and KLOE experiments had ever searched for the dark photon in $e^+e^-$ collision \cite{Lees:2014xha,Ablikim:2017aab,Anastasi:2015qla}. However, the mass range in BaBar experiment was in between $20$ MeV $\sim 10.2$ \text{GeV}, and BESIII searched for the dark photon in the $1.5\sim3.4$ GeV mass region, both overshot the 16.7 MeV $X$ boson. The CLOE experiment explored the 5$\sim$520 MeV range, yet did not find clear signatures of the dark photon, whereas it constrained the parameter $g^v_e$ to be $|g^v_e|\leq2\times 10^{-3}$, which is in accordance with what we employ in this work.

Note that, other proposals of searching for an extra U(1) gauge boson $U$ with its decays into $e^+e^-$ or invisible at electron-positron colliders are also provided in Ref. \cite{Araki:2017wyg,Alikhanov:2017cpy}, as well as the background analysis \cite{Kozhuharov:2017qjo}, yet with different methods/focus in comparison with the present work.
\begin{center}
\begin{table}[t]
\caption{The signal-to-noise ratios (SNR) around $M_{ee}=16.7$ MeV, where we adopt four invariant-mass ranges $M_{ee}$ for the background process. The last three ranges roughly correspond to the energy resolution of $\delta_E=2,~4,~5$ MeV (Fig. \ref{smear} (b,~c,~d)) respectively. Note, signal events would be suppressed by two orders of magnitude when taking $g^{v/a}_e=10^{-4}$.}\label{snr}
\begin{tabular}{c| c| c| c| c  }
\hline
$M_{ee}(\sqrt{s_{ee}})$ Ranges (MeV) & [16.6, 16.8] & [10, 25] & [5, 30] & [2, 33] \\ \hline
Background Events (B) & 4961 & $3.80\times10^5$ & $7.42\times10^5$ & $1.16\times10^6$\\ \hline
Signal Events (S) & \multicolumn{4}{c}{5952 ($\sqrt{s}=3.7~\text{GeV},~g^{v/a}_e=10^{-3}$)}\\ \hline
SNR ($\frac{S}{\sqrt{S+B}}$) & 57 & 9.6 & 6.9 & 5.5 \\
\hline
\end{tabular}
\end{table}
\end{center}

\subsection{$J/\psi \to X+\gamma$ with $X \to e^+e^-$}

Up to date the $J/\psi$ events collected at BESIII are $(1310.6\pm7.0)\times10^6$ \cite{Ablikim:2016fal}, which is quite abundant. Hence in this subsection, we will consider the $X(16.7)$ boson production in $J/\psi$ decays seriously.

\begin{figure}[t]
\begin{center}
\includegraphics[scale=0.46]{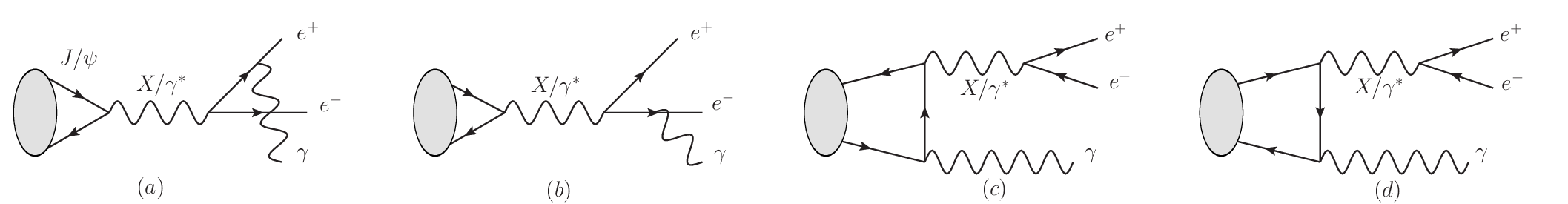}
\caption{The Feynman diagrmas for the signal/background processes $J/\psi \xrightarrow[]{X/\gamma^*} e^+e^-\gamma$.}\label{psig}
\end{center}
\end{figure}

In this subsection, we continue to take the Lagrangian terms as shown in Eq. \ref{lag}, \emph{i.e.} the interaction vertex is still the ``vector minus axial-vector (V-A)'' type. The Feynman diagrams can be easily obtained from Fig. \ref{psig} (c,d), where the $X$ boson decays into the $e^+e^-$ pairs. Then the decay width reads
\bqa
\Gamma(J/\psi\to X\gamma)=\frac{8 \pi \alpha ^2 (g^a_c)^2 \Psi^2 (16 m_c^4+40 m_c^2 m_X^2+m_X^4)}{27 m_c^4 (4 m_c^2-m_X^2)},
\eqa
where the squared wave function at the origin $\Psi^2=\frac{(2m_c)^2\Gamma(J/\psi\to e^+e^-)}{16\pi e_c^2 \alpha^2}$ \cite{Braaten:2002fi}, with $e_c=2/3$ and $\Gamma(J/\psi\to e^+e^-)=5.55\times10^{-6}$ GeV \cite{Patrignani:2016xqp}. Obviously, the contribution of vector current (``V only'') vanishes.\footnote{Landau-Yang theorem: Spin-1 particles cannot decay into two photons, which is forbidden by the conservation of orbital angular momentum.} Taking the total decay width of $J/\psi$ as 92.9 keV, then we obtain the branching fraction, $0.0398\times (g_c^a)^2$. Given the $1.3\times10^9$ $J/\psi$ events and $g_c^a=10^{-4}\sim10^{-3}$, the expected events in $J/\psi\to X\gamma$ would be $1(0.52)\sim52$ (values in brackets are the estimate before rounding up).

The $X$ boson signals will also be captured through $X \to e^+e^-$ decay in experiments. Feynman diagrams of the related processes $J/\psi \xrightarrow[]{X/\gamma^*} e^+e^-\gamma$ are presented in Fig. \ref{psig}.  One may notice that if $X$ boson was a massive $\gamma$-like particle, there would be no contributions from the last two figures of Fig. \ref{psig} (c,~d). And we also find that the contribution of axial-vector current (``A only'') from Fig. \ref{psig} (a,~b) vanishes at tree amplitudes level. According to the numerical evaluation (assuming $g_c^{v/a}= g_e^{v/a}=10^{-3}$), it is found that contribution of vector current (``V only'') of Fig. \ref{psig} (a,~b) is suppressed by a factor of $10^{-12}$ in comparison with that of the background process $J/\psi \xrightarrow[]{\gamma^*} e^+e^-\gamma$. So it can be concluded that the main contributions to the decay width of $J/\psi \xrightarrow[]{X} e^+e^-\gamma$ come from the axial-vector current interaction in Fig. \ref{psig} (c,~d).

\begin{figure}[t]
\begin{center}
\includegraphics[scale=0.56]{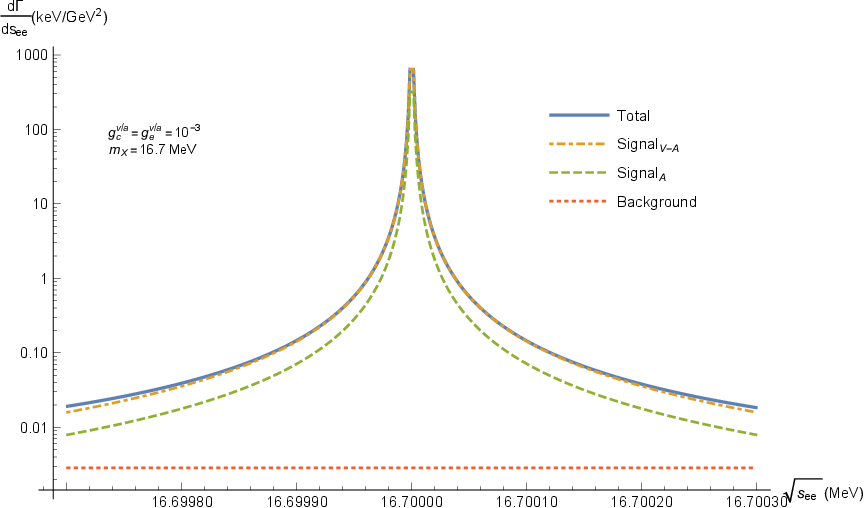}
\includegraphics[scale=0.25]{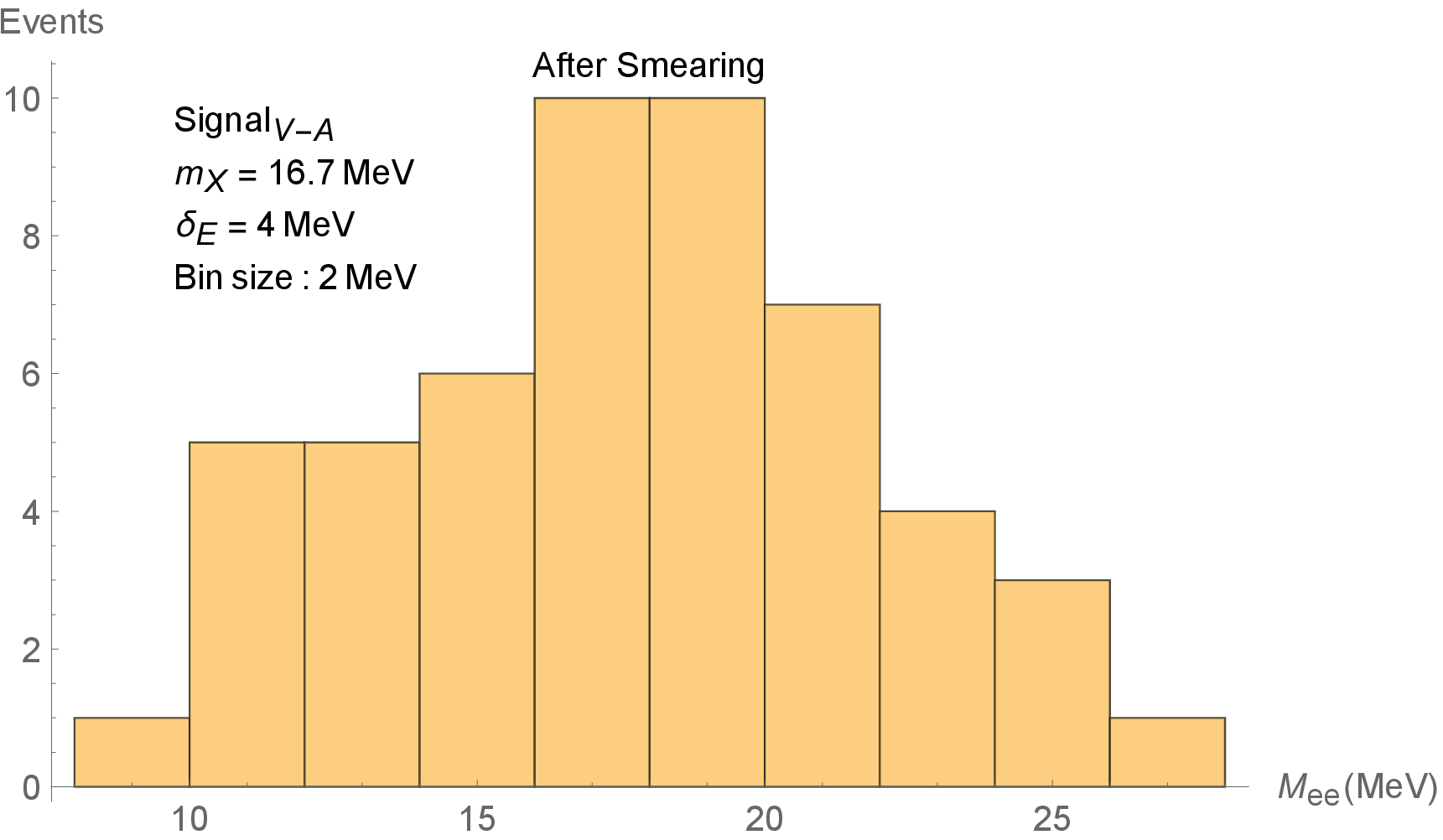}
\caption{Left: the differential decay-width with respect to the invariant-mass of $e^+e^-$ pairs for the $J/\psi \xrightarrow[]{X/\gamma^*} e^+e^-\gamma$ processes. Right: the resonance bump of ``Signal$_{V-A}$'' simulated
with 52 $X$ events after smearing with the energy resolution of $\delta_E=4$ MeV.}\label{psiee}
\end{center}
\end{figure}

The invariant-mass distribution of electron-positron pairs in final states of processes $J/\psi \xrightarrow[]{X/\gamma^*} e^+e^-\gamma$ has been presented in Fig. \ref{psiee} (left), where we adopt no cuts on emitting polar angles of the final particles. Obviously, the ``Signal$_{V-A}$'' line shape overlaps with the ``Total'' one, which implies that the contribution of cross-terms between the $X$-propagated Feynman diagrams and the $\gamma^*$-propagated ones is negligible. And we find that around the $X$ resonance, the differential decay widths of the signals are much larger than the ones of background, \emph{i.e.}, the signal to noise ratio is dramatic before smearing. Here, we evaluate the contribution from the axial-vector current purposely (the ``Signal$_{A}$'' dashed line), \emph{i.e.}, the contribution of the axial-vector current in Fig. \ref{psig} (c,~d). And it is about one half of the contribution of ``Signal$_{V-A}$'', out of our expectation discussed in above paragraph. We find that the other half also comes from the Fig. \ref{psig} (c,~d), which is the contribution of the cross-term between the axial-vector term $i g_c^a\gamma_{\mu}\gamma_5$ in $X-c\bar{c}$ vertex and the vector term $i g_e^v\gamma_{\mu}$ in $X-e^+e^-$ vertex, which has been excluded in the ``Signal$_{A}$'' case.

In Fig. \ref{psiee} (right), we present the distribution of the estimated 52 $X$ boson events in the $M_{ee}$ spectrum after the smearing with the energy resolution of $\delta_E=4$ MeV. In Tab. \ref{snr2}, we have considered several $M_{ee}$ ranges around $X$ resonance, and the related signal-to-noise ratios (SNR) are presented. Note that we adopt no cuts on the emitting polar angles of the final particles here. And the third range is roughly corresponding to the energy resolution of $\delta_E=4$ MeV (Fig. \ref{psiee} (right)). Given the coupling strength $\sim10^{-3}$, although we have only 52 $X$ events, the amount of the background events are even fewer comparatively, which implies that there might be some good news of trapping the $X$ boson in $J/\psi \to e^+e^-\gamma$ process.

In fact, the BESIII collaboration once searched for a light exotic particle $A^0$ in $\psi^{\prime} \to J/\psi\pi^+\pi^-, J/\psi \to A^0\gamma, A^0\to\mu^+\mu^-$, and found one event with the $\mu^+\mu^-$ mass of 213.3 MeV using a sample of $1.06\times10^8~\psi^{\prime}$ \cite{Ablikim:2011es}. Recently, they searched for $A^0$ again in the mass range of 0.212$\sim$3.0 GeV using $(225.0\pm2.8)\times10^6$ $J/\psi$ events, found no evidence but upper limits on the product branching fraction of $J/\psi \to A^0\gamma, A^0\to\mu^+\mu^-$ \cite{Ablikim:2015voa}. Here, we strongly suggest searching for the light axial-vector $X(16.7)$ resonance in the $e^+e^-$ invariant-mass spectrum using the substantial $J/\psi$ data collected in BESIII.

Note that, in contrast with the vector current coupling, such $X$ boson accounting for the isoscalar $^8\text{Be}^{*} \to$ $^8\text{Be}$ transition anomaly with its couplings to quarks and leptons through an axial-vector current provides a natural suppression of the isovector $^8\text{Be}^{*\prime} \to$ $^8\text{Be}$ transition \cite{Kozaczuk:2016nma}. Moreover, a theory with axial-vector couplings motivated by several MeV-scale anomalies can be UV-completed consistent with the Standard Model gauge invariance, see Ref.s \cite{Kahn:2016vjr,Ismail:2016tod} for details.
In Ref. \cite{Kozaczuk:2016nma}, according to their simple UV-complete model, the axial-vector couplings to $u,d$ quarks were constrained to be at the order of $10^{-5}\sim10^{-4}$, but this still leaves room to have a large enough coupling to explain the $^8\text{Be}$ anomaly, say if the assumptions about the couplings they adopted were relaxed as mentioned there, or in other consistent UV models. In the minimal flavor violating limit, one may assume the charm axial-vector coupling to be same as the up quark one. In this case, and consider the $X-c\bar{c}$ axial-vector coupling to be at the order of $10^{-4}$, the expected $X(16.7)$ events in $J/\psi$ decays would be about 1 as estimated above. This implies that the $J/\psi$ search for the $X(16.7)$ boson at BESIII will not reach the sensitivity below $10^{-4}$ that Ref. \cite{Kozaczuk:2016nma} suggests, with the $1.3\times10^9~J/\psi$ events collected to date.
\begin{center}
\begin{table}[h]
\caption{The signal-to-noise ratios (SNR) around the $X$ resonance mass (16.7 MeV), where we adopt four invariant-mass ranges for the background process. Note, signal events would be suppressed by two orders of magnitude when taking $g^{a}_{c}=10^{-4}$.}\label{snr2}
\begin{tabular}{c| c| c| c| c }
\hline
 $M_{ee}(\sqrt{s_{ee}})$ Ranges (MeV) & [16.6, 16.8] & [10, 25] & [5, 30] & [2, 33] \\ \hline
Background Events (B) & 0 & 22 & 37 & 46\\ \hline
Signal Events (S) & \multicolumn{4}{c}{52 ($g^{a}_{c} = 10^{-3}$)}\\ \hline
SNR ($\frac{S}{\sqrt{S+B}}$) & 7.2 & 6.1 & 5.5 & 5.2 \\
\hline
\end{tabular}
\end{table}
\end{center}

\section{Summary and Prospect}

In summary, inspired by the 6.8$\sigma$ anomaly in $^8\text{Be}$ nuclear transition experiment and the passion of searching for the new gauge boson, we investigate the possibility of hunting for this yet-not-verified gauge boson $X(16.7)$ in both $e^+e^-$ collision and $J/\psi$ decays. In our model, the complete ``vector minus axial-vector'' interaction vertex is taken into account. We have set two traps for the $X(16.7)$ hunting: $e^+e^- \to X\gamma$ and $J/\psi \to X\gamma$, both following with the saturated $X\to e^+e^-$ decay. Analytical results of differential distribution in $e^+e^- \to X\gamma$ and the decay widths of both $J/\psi \to X\gamma$ and $X \to e^+e^-$ are presented and discussed. Phenomenologically, we evaluate the decay length of $X$ boson in the experiment frame, the production rates, the cross-section/decay-width differential distribution with respect to the $e^+e^-$ invariant-mass spectrum, and particularly the signal-to-noise ratios, which may be increased by the $X$ boson decay length, before and after the smearing for both processes.

For $X$ boson production in $e^+e^-$ collision at $\sqrt{s}=3.7$ GeV, given the coupling strength $g_e^{v/a}=10^{-4}\sim10^{-3}$, it is found that the decay length is $0.27$ mm $ < L < 27$ \text{mm}, the expected $X$ boson events are $60 \sim 6000$ per year, and the signal-to-noise ratios (SNR) decreases  significantly after smearing. We also find that contributions of vector and axial-vector currents are of the equal importance. While in $J/\psi$ decays, we find that the axial-vector current dominates the $X(16.7)$ production associated with a photon. Given copious $1.3\times10^9~J/\psi$ events at BESIII and $g_{c}^{a}= 10^{-3}$, there would be only 52 $X$ events within the data, while the SNR are inspiring after smearing. That is, though $e^+e^-\to X\gamma$ process may yield more signals, it has a relatively larger background in regard to $J/\psi \to X\gamma$ process.

In all, by measuring the final state $e^+e^-$ invariant-mass spectrum, we find it is possible for BESIII experiment to perform a decisive measurement on the $X(16.7)$ boson.

\vspace{.7cm}
{\bf Acknowledgments}

We thank Pei-Rong Li, Xiao-Rui Lyu and Da-Yong Wang for useful discussions on the $X$ detection at BESIII, and appreciate very much the anonymous reviewer's valuable comments and suggestions. This work was supported in part by the Ministry of Science and Technology of the Peoples' Republic of China(2015CB856703); by the Strategic Priority Research Program of the Chinese Academy of Sciences, Grant No.XDB23030100; by the National Natural Science Foundation of China(NSFC) under the grants 11375200 and 11635009;
and by the China Postdoctoral Science Foundation (Grant No. 2017M610973).




\begin{thebibliography}{99}

\bibitem{Patrignani:2016xqp}
  C.~Patrignani {\it et al.} [Particle Data Group],
  Chin.\ Phys.\ C {\bf 40}, no. 10, 100001 (2016).

\bibitem{Aad:2015osa}
  G.~Aad {\it et al.} [ATLAS Collaboration],
  JHEP {\bf 1507}, 157 (2015),
  [arXiv:1502.07177 [hep-ex]].

\bibitem{CMS:2016zxk}
  CMS Collaboration [CMS Collaboration],
  CMS-PAS-EXO-16-008.

\bibitem{ATLAS:2016cyf}
  The ATLAS collaboration [ATLAS Collaboration],
  ATLAS-CONF-2016-045.

\bibitem{CMS:2016abv}
  CMS Collaboration [CMS Collaboration],
  CMS-PAS-EXO-16-031.

\bibitem{Policicchio:2016evl}
  A.~Policicchio [ATLAS Collaboration],
  PoS ICHEP {\bf 2016}, 1149 (2017).

\bibitem{Lees:2014xha}
  J.~P.~Lees {\it et al.} [BaBar Collaboration],
  Phys.\ Rev.\ Lett.\  {\bf 113}, no. 20, 201801 (2014),
  [arXiv:1406.2980 [hep-ex]].

\bibitem{Ablikim:2017aab}
  M.~Ablikim {\it et al.} [BESIII Collaboration],
  Phys.\ Lett.\ B {\bf 774}, 252 (2017),
  [arXiv:1705.04265 [hep-ex]].

\bibitem{Anastasi:2015qla}
  A.~Anastasi {\it et al.},
  Phys.\ Lett.\ B {\bf 750}, 633 (2015),
  [arXiv:1509.00740 [hep-ex]].

\bibitem{Babusci:2012cr}
  D.~Babusci {\it et al.} [KLOE-2 Collaboration],
  Phys.\ Lett.\ B {\bf 720}, 111 (2013),
  [arXiv:1210.3927 [hep-ex]].

\bibitem{Batley:2015lha}
  J.~R.~Batley {\it et al.} [NA48/2 Collaboration],
  Phys.\ Lett.\ B {\bf 746}, 178 (2015),
  [arXiv:1504.00607 [hep-ex]].

\bibitem{Agakishiev:2013fwl}
  G.~Agakishiev {\it et al.} [HADES Collaboration],
  Phys.\ Lett.\ B {\bf 731}, 265 (2014),
  [arXiv:1311.0216 [hep-ex]].

\bibitem{Erler:2009jh}
  J.~Erler, P.~Langacker, S.~Munir and E.~Rojas,
  JHEP {\bf 0908}, 017 (2009),
  [arXiv:0906.2435 [hep-ph]].

\bibitem{Bilmis:2015lja}
  S.~Bilmis, I.~Turan, T.~M.~Aliev, M.~Deniz, L.~Singh and H.~T.~Wong,
  Phys.\ Rev.\ D {\bf 92}, no. 3, 033009 (2015),
  [arXiv:1502.07763 [hep-ph]].

\bibitem{Ilten:2015hya}
  P.~Ilten, J.~Thaler, M.~Williams and W.~Xue,
  Phys.\ Rev.\ D {\bf 92}, no. 11, 115017 (2015),
  [arXiv:1509.06765 [hep-ph]].

\bibitem{Wojtsekhowski:2017ijn}
  B.~Wojtsekhowski {\it et al.},
  arXiv:1708.07901 [hep-ex].

\bibitem{Corliss:2017tms}
  R.~Corliss [DarkLight Collaboration],
  Nucl.\ Instrum.\ Meth.\ A {\bf 865}, 125 (2017).

\bibitem{Gninenko:2017acc}
  S.~N.~Gninenko and N.~V.~Krasnikov,
  doi:10.3204/DESY-PROC-2016-04/Krasnikov

\bibitem{Alexander:2016aln}
  J.~Alexander {\it et al.},
  arXiv:1608.08632 [hep-ph].

\bibitem{Kozhuharov:2016tdb}
  V.~Kozhuharov, M.~Raggi and P.~Valente,
  arXiv:1610.04389 [hep-ex].

\bibitem{Denig:2016dgi}
  A.~Denig,
  EPJ Web Conf.\  {\bf 130}, 01005 (2016).

\bibitem{Krasznahorkay:2015iga}
  A.~J.~Krasznahorkay {\it et al.},
  Phys.\ Rev.\ Lett.\  {\bf 116}, no. 4, 042501 (2016),
  [arXiv:1504.01527 [nucl-ex]].

\bibitem{Krasznahorkay:2017bwh}
  A.~J.~Krasznahorkay {\it et al.},
  PoS BORMIO {\bf 2017}, 036 (2017).

\bibitem{Krasznahorkay:2017gwn}
  A.~J.~Krasznahorkay {\it et al.},
  EPJ Web Conf.\  {\bf 142}, 01019 (2017).

\bibitem{Krasznahorkay:2017qfd}
  A.~J.~Krasznahorkay {\it et al.},
  EPJ Web Conf.\  {\bf 137}, 08010 (2017).

\bibitem{Feng:2016jff}
  J.~L.~Feng, B.~Fornal, I.~Galon, S.~Gardner, J.~Smolinsky, T.~M.~P.~Tait and P.~Tanedo,
  Phys.\ Rev.\ Lett.\  {\bf 117}, no. 7, 071803 (2016),
  [arXiv:1604.07411 [hep-ph]].

\bibitem{Gu:2016ege}
  P.~H.~Gu and X.~G.~He,
  Nucl.\ Phys.\ B {\bf 919}, 209 (2017),
  [arXiv:1606.05171 [hep-ph]].

\bibitem{Ellwanger:2016wfe}
  U.~Ellwanger and S.~Moretti,
  JHEP {\bf 1611}, 039 (2016),
  [arXiv:1609.01669 [hep-ph]].

\bibitem{Feng:2016ysn}
  J.~L.~Feng, B.~Fornal, I.~Galon, S.~Gardner, J.~Smolinsky, T.~M.~P.~Tait and P.~Tanedo,
  Phys.\ Rev.\ D {\bf 95}, no. 3, 035017 (2017)
  [arXiv:1608.03591 [hep-ph]].

\bibitem{Kozaczuk:2016nma}
  J.~Kozaczuk, D.~E.~Morrissey and S.~R.~Stroberg,
  Phys.\ Rev.\ D {\bf 95}, no. 11, 115024 (2017),
  [arXiv:1612.01525 [hep-ph]].

\bibitem{Neves:2016ugb}
  M.~J.~Neves and J.~A.~Helay$\ddot{e}$l-Neto,
  arXiv:1611.07974 [hep-ph].

\bibitem{DelleRose:2017phz}
  L.~Delle Rose, S.~Khalil and S.~Moretti,
  arXiv:1708.08806 [hep-ph].

\bibitem{Neves:2017rcn}
  M.~J.~Neves,
  arXiv:1704.02491 [hep-ph].

\bibitem{Fornal:2017msy}
  B.~Fornal,
  Int.\ J.\ Mod.\ Phys.\ A {\bf 32}, 1730020 (2017)
  [arXiv:1707.09749 [hep-ph]].

\bibitem{Zhang:2017zap}
  X.~Zhang and G.~A.~Miller,
  Phys.\ Lett.\ B {\bf 773}, 159 (2017),
  [arXiv:1703.04588 [nucl-th]].

\bibitem{Kahn:2016vjr}
  Y.~Kahn, G.~Krnjaic, S.~Mishra-Sharma and T.~M.~P.~Tait,
  JHEP {\bf 1705}, 002 (2017),
  [arXiv:1609.09072 [hep-ph]].

\bibitem{Ismail:2016tod}
  A.~Ismail, W.~Y.~Keung, K.~H.~Tsao and J.~Unwin,
  Nucl.\ Phys.\ B {\bf 918}, 220 (2017),
  [arXiv:1609.02188 [hep-ph]].

\bibitem{Kitahara:2016zyb}
  T.~Kitahara and Y.~Yamamoto,
  Phys.\ Rev.\ D {\bf 95}, no. 1, 015008 (2017),
  [arXiv:1609.01605 [hep-ph]].

\bibitem{Jia:2016uxs}
  L.~B.~Jia and X.~Q.~Li,
  Eur.\ Phys.\ J.\ C {\bf 76}, no. 12, 706 (2016),
  [arXiv:1608.05443 [hep-ph]].

\bibitem{Chen:2016tdz}
  C.~S.~Chen, G.~L.~Lin, Y.~H.~Lin and F.~Xu,
  Int.\ J.\ Mod.\ Phys.\ A {\bf 32} (2017) no.31, 1750178,
  [arXiv:1609.07198 [hep-ph]].

\bibitem{Yamamoto:2017ypv}
  Y.~Yamamoto,
  EPJ Web Conf.\  {\bf 168}, 06007 (2018),
  [arXiv:1708.09756 [hep-ph]].

\bibitem{Liang:2016ffe}
  Y.~Liang, L.~B.~Chen and C.~F.~Qiao,
  Chin.\ Phys.\ C {\bf 41}, no. 6, 063105 (2017),
  [arXiv:1607.08309 [hep-ph]].

\bibitem{Jia:2017iyc}
  L.~B.~Jia,
  arXiv:1710.03906 [hep-ph].

\bibitem{Chen:2016kxw}
  C.~H.~Chen and T.~Nomura,
  Phys.\ Lett.\ B {\bf 763}, 304 (2016),
  [arXiv:1608.02311 [hep-ph]].

\bibitem{Chiang:2016cyf}
  C.~W.~Chiang and P.~Y.~Tseng,
  Phys.\ Lett.\ B {\bf 767}, 289 (2017),
  [arXiv:1612.06985 [hep-ph]].

\bibitem{Raggi:2014zpa}
  M.~Raggi and V.~Kozhuharov,
  Adv.\ High Energy Phys.\  {\bf 2014}, 959802 (2014),
  [arXiv:1403.3041 [physics.ins-det]].

\bibitem{Araki:2017wyg}
  T.~Araki, S.~Hoshino, T.~Ota, J.~Sato and T.~Shimomura,
  Phys.\ Rev.\ D {\bf 95}, no. 5, 055006 (2017),
  [arXiv:1702.01497 [hep-ph]].

\bibitem{Kozaczuk:2017per}
  J.~Kozaczuk,
  arXiv:1708.06349 [hep-ph].

\bibitem{Alikhanov:2017cpy}
  I.~Alikhanov and E.~A.~Paschos,
  arXiv:1710.10131 [hep-ph].

\bibitem{Kozhuharov:2017qjo}
  V.~Kozhuharov,
  EPJ Web Conf.\  {\bf 142}, 01018 (2017).


\bibitem{peskin} M. E. Peskin and D. V. Schroeder, {\it An Introduction to Quantum Field Theory}, Westview Press, 1995. Page 168.

\bibitem{Han:2005mu}
  T.~Han,
  hep-ph/0508097.

\bibitem{Aubert:2001tu}
  B.~Aubert {\it et al.} [BaBar Collaboration],
  Nucl.\ Instrum.\ Meth.\ A {\bf 479}, 1 (2002),
  [hep-ex/0105044].

\bibitem{Pospelov:2008zw}
  M.~Pospelov,
  Phys.\ Rev.\ D {\bf 80}, 095002 (2009),
  [arXiv:0811.1030 [hep-ph]].

\bibitem{Ablikim:2009aa}
  M.~Ablikim {\it et al.} [BESIII Collaboration],
  Nucl.\ Instrum.\ Meth.\ A {\bf 614}, 345 (2010),
  [arXiv:0911.4960 [physics.ins-det]].


\bibitem{Ablikim:2016fal}
  M.~Ablikim {\it et al.} [BESIII Collaboration],
  Chin.\ Phys.\ C {\bf 41}, no. 1, 013001 (2017),
  [arXiv:1607.00738 [hep-ex]].

\bibitem{Braaten:2002fi}
  E.~Braaten and J.~Lee,
  Phys.\ Rev.\ D {\bf 67}, 054007 (2003),
  Erratum: [Phys.\ Rev.\ D {\bf 72}, 099901 (2005)],
  [hep-ph/0211085].

\bibitem{Ablikim:2011es}
  M.~Ablikim {\it et al.} [BESIII Collaboration],
  Phys.\ Rev.\ D {\bf 85}, 092012 (2012),
  [arXiv:1111.2112 [hep-ex]].

\bibitem{Ablikim:2015voa}
  M.~Ablikim [BESIII Collaboration],
  Phys.\ Rev.\ D {\bf 93}, no. 5, 052005 (2016),
  [arXiv:1510.01641 [hep-ex]].


\end{thebibliography}
\end{document}